\documentclass[fleqn,11pt]{article}
\usepackage[dvips]{graphicx}

\author{ M. G\'omez-Bock, and  R. Noriega-Papaqui\\
{\small \it Instituto de F\'{\i}sica, Benem\'erita Universidad
Aut\'onoma de Puebla},\\
{\small \it Apartado Postal J-48 Puebla, Pue. M\'{e}xico}}

\title{Flavor violating decays of the Higgs bosons in the THDM-III}
\begin{document}
\maketitle


\begin{abstract}
 We calculate the branching ratios for the decays of neutral Higgs bosons
($h^{0},H^{0},A^{0}$) into pairs of fermions, including flavor
violating processes, in the context of the General Two Higgs
Doublet Model III.
\end{abstract}

\section{Introduction}
The mechanism that generates the fermion masses is perhaps hidden
behind the structure of the Yukawa couplings. In the Standard
Model (SM) the form of these couplings is not well understood and
neither are their origin nor the underlying principles (the flavor
problem) \cite{FN}, \cite{cheng-li}, \cite {Fritzsch:2001py}.
Phenomenologically, the SM parameterizes the values of the Yukawa
couplings, but, theoretically, we are in the darkness. This is one
of the reasons why many physicists agree that the SM should be
considered as an effective theory that remains valid up to an
energy scale of \textit{O}(TeV), and eventually will be replaced
by a more fundamental theory. A way to picture this could be to
consider that the SM is the top of an iceberg, so that a great
deal still remains to be explored, in order to understand what
lies underneath.

One simple extension of the SM adds a new Higgs doublet, and it is
known as the Two Higgs Doublet Model (THDM)\cite{glashow},
\cite{higgshunter}; which constitutes precisely the structure that
is required in the construction of the Minimal Supersymmetric SM
(MSSM) \cite{Martin:1997ns}. The direct consequences of this
extension are: an increase in the scalar spectrum, a more generic
pattern of Flavor Changing Neutral Currents (FCNC), including FCNC
at tree level, which are highly excluded in low energy experiments
and turn out to be a potential problem. In the earlier versions
(THDM I-II), this problem was solved by imposing a discrete
symmetry that restricts each fermion to be coupled at most to one
Higgs
doublet,\cite{glashow}, \cite{Savage:1991qh}.\\

In the so called version III of the model, FCNC are kept under
control by imposing a certain form for the Yukawa matrices that
reproduce the observed fermion masses and mixing angles
\cite{Cheng:1987rs}. This more general model allows a rich
phenomenology through extra terms that come from the Yukawa
couplings, and that is present even at tree level. The use of {\it
texture forms}\cite{Fritzsch:1999ee} permits to establish a direct
relation from the elements of the matrix with the mixing parameter
used in calculating the branching ratios, without dropping terms
proportional to the lighter fermion masses in advanced.
Specifically, considering an hermitian Yukawa matrix of the {\it
6-textures type} one gets the {\it Cheng-Sher Ansatz} (the flavor
violation couplings are given as proportional to the square root
of the mass product) for the flavor mixing couplings which is
widely used in literature \cite{ansatz}.\\

In this work, we departure from a {\it 4-texture form} of the
Yukawa matrices. We are interested in employing this version
(THDM-III) to evaluate the branching ratios of the neutral Higgs
bosons decaying into fermion pairs, with the idea of studying the
model predictions and its test at future Large Hadron Collider (LHC) data.\\

Hence, considering that the flavor violating interactions arise
from the Yukawa couplings, our starting point is an {\it Hermitian
4-texture ansatz} in order to construct the mass matrices, which
have been found to be in agreement with the observed data
\cite{fourtext},\cite{Xing}. The relation from the matrix elements
to the mixing parameter used to calculate FCNC processes in the
Higgs sector is developed in \cite{primero}. Some Higgs
phenomenology have been addressed in the literature, mainly
aimed to describe the light Higgs boson \cite{segundo}, \cite{Alexis}.\\

The paper is organized as follows. In Sec. II we discuss the
structure of the Yukawa lagrangian of the THDM-III, and present
the couplings $\phi^{0} f_{i}f_{j}$, ($\phi^{0}=
h^{0},H^{0},A^{0}$) by using an hermitian four-texture form for
the mass matrices; we also show how the Yukawa couplings induce
the LFV Higgs decays. Then, in Sec.III we present the analytical
formulae for the couplings of the three neutral Higgs bosons with
fermions and we compare them with the MSSM and for the flavor
conserving modes with those of the SM case. We also present the
formulae used for the calculations of the branching ratios. The
results are presented in Sec. IV and some conclusions are left for Sec. V.\\

\section{Structure of the Yukawa sector in the  Two Higgs Doublet Model Type III}

The couplings of the Higgs boson to the fermions are given by the
Yukawa lagrangian, which in general can be written  as follows

\begin{equation}
L_{Y}=\sum_{a,i}Y_{a}^{i} \overline{F}_{L}^{i}\Phi_a
f_{R}^{i}+h.c.,
\end{equation}
where $F_{L}$ denotes the fermion doublet (left-handed), $f_{R}$
is the fermion singlet (right-handed), and $\Phi_a$ are the two
Higgs doublets $(a=1,2)$. Considering the three generations, the
coefficient $Y_{a}^{i}$ can be expressed as a $3\times3$ matrix
and tagged as $Y_{a}^{l}, Y_{a}^{u}, Y_{a}^{d}$, for leptons, $u$
and $d$ type quarks. Here we consider massless neutrinos.\\

What we know so far about the Yukawa terms is that they should
reproduce the fermion masses, the mixing angles of the CKM matrix
and keep the FCNC under control, as they are restricted by
experimental data. From the form of the lagrangian, it is natural
to consider that the flavor violation or the mixtures between
families, could arise directly from the form of the Yukawa terms,
which, in general, are not diagonal. In this work we are
interested in the study of the Higgs boson decays $H \to
f_{i}f_{j}$, as a possible signal of fermion flavor violation for
the complete spectrum of neutral Higgs bosons $h^{0},H^{0},A^{0}$ of the THDM-III.\\

In particular, in the Yukawa sector for the THDM-III, both Higgs
doublets may couple with the two types of fermions, \textit{i.e.,
up} and \textit{down}, so that we have two different Yukawa terms
for each doublet, $Y_a$ with $a=1,2$. Thus, the Yukawa lagrangian
is given by:
\begin{equation}
{\cal{L}}_Y^l = Y^{l}_{1}\bar{L'_{L}}\Phi_{1}l'_{R} +
Y^{l}_{2}\bar{L'_{L}}\Phi_{2}l'_{R};
\label{lagleptones}
\end{equation}

\begin{equation}
{\cal{L}}_Y^q = Y^{u}_1\bar{Q'}_L {\tilde \Phi_{1}} u'_{R} +
                   Y^{u}_2 \bar{Q'}_L {\tilde \Phi_{2}} u'_{R} +
Y^{d}_1\bar{Q'}_L \Phi_{1} d'_{R} + Y^{d}_2
\bar{Q'}_L\Phi_{2}d'_{R}, \label{lagquarks}
\end{equation}
\noindent where the first equation corresponds to the leptonic
sector and the second one to the quark sector. $\Phi_{1,2}$ refer
to the two Higgs doublets, where ${\tilde \Phi_{i}}=i \sigma_{2}\Phi_{i}^* $
and $Y_{1,2}^{u,d,l}$ denote the $(3 \times 3)$ Yukawa matrices.\\

After SSB (Spontaneous Symmetry Breaking), one can derive the
fermion mass matrices from eqs. (\ref{lagleptones}) and
(\ref{lagquarks}), namely \\
\begin{equation}
M_f= \frac{1}{\sqrt{2}}(v_{1}Y_{1}^{f}+v_{2}Y_{2}^{f}), \qquad
f = u, d, l,
\label{masa-fermiones}
\end{equation}
\noindent Here, we are taking into account the fact that, working
with a \textit{hierarchical ansatz} for the mass matrix and by
means of equation (\ref{masa-fermiones}), the simplest case is to
consider that both Yukawas $Y^f_{1,2}$ possess the same structure
(without anomalous cancellation of any of the elements of the
matrices), particularly we use an hermitian 4-texture form, and
because of eq. (\ref{masa-fermiones}) the complete mass matrix
inherits this structure. The mass matrix is diagonalized through
the bi-unitary matrices $V_{L,R}$, though each Yukawa matrices are
not diagonalized by this transformation. The diagonalization is
performed in the following way

\begin{equation}
\bar{M}_f = V_{fL}^{\dagger}M_{f}V_{fR}. \label{masa-diagonal}
\end{equation}

The fact that $M_f$ is hermitian, under the considerations given
above (hermitian Yukawa matrices), directly implies that $V_{fL} =
V_{fR}$, and the mass eigenstates for the fermions are given by

\begin{equation}
u = V_{u}^{\dagger}u' \qquad d = V_{d}^{\dagger}d' \qquad l =
V_{l}^{\dagger}l'. \label{redfields}
\end{equation}

Then eq. (\ref{masa-fermiones}) in this basis takes the form
\begin{equation}
\bar{M}_f=\frac{1}{\sqrt{2}}(v_{1}\tilde{Y}_{1}^{f}+v_{2}
\tilde{Y}_{2}^{f})
\end{equation}
where $\tilde{Y}_{i}^{f}=V_{fL}^{\dagger}Y_{i}^{f}V_{fR}$ and for
the quark case we may write
\begin{eqnarray}
\tilde{Y}_{1}^{d}&=&
\frac{\sqrt{2}}{v\cos\beta}\bar{M}_{d}-\tan\beta\tilde{Y}_{2}^{d}\nonumber \\
\tilde{Y}_{2}^{u}&=&\frac{\sqrt{2}}{v\sin\beta}\bar{M}_{u}-\cot\beta\tilde{Y}_{1}^{u}
\label{rotyukawas}
\end{eqnarray}
In the lepton case we perform the usual substitution $d \rightarrow l$.\\

By using the redefined fields eq. (\ref{redfields}), as the
physical states, and considering the Yukawas in this basis, eq.
(\ref{rotyukawas}), we rewrite the THDM-III Yukawa Lagrangian in
terms of the mass eigenstates. The interactions of the neutral
Higgs bosons $(h^{0}, H^{0}, A^{0})$ with quark pairs acquire the
following
form:\\

\begin{eqnarray}
{\cal{L}}_Y^{q} & = &
\frac{g}{2}\left(\frac{m_{d_{i}}}{m_W}\right)
\bar{d_{i}}\left[\frac{ \, \cos\alpha}{\cos\beta}\delta_{ij}+
\frac{\sqrt{2} \, \sin(\alpha - \beta)}{g \, \cos\beta}
\left(\frac{m_W}{m_{d_{i}}}\right)(\tilde{Y}_2^d)_{ij}\right]d_{j}H^{0}
\nonumber \\
&  &+ \frac{g}{2}\left(\frac{m_{d_{i}}}{m_W}\right)\bar{d}_{i}
\left[-\frac{\sin\alpha}{\cos\beta} \delta_{ij}+ \frac{\sqrt{2} \,
\cos(\alpha - \beta)}{g \, \cos\beta}
\left(\frac{m_W}{m_{d_{i}}}\right)(\tilde{Y}_2^d)_{ij}\right]d_{j}
h^{0}
\nonumber \\
& &+ \frac{ig}{2}\left(\frac{m_{d_{i}}}{m_W}\right)\bar{d}_{i}
\left[-\tan\beta \delta_{ij}+  \frac{\sqrt{2} }{g \, \cos\beta}
\left(\frac{m_W}{m_{d_{i}}}\right)(\tilde{Y}_2^d)_{ij}\right]
\gamma^{5}d_{j} A^{0} \nonumber \\
& &+ \frac{g}{2}\left(\frac{m_{u{i}}}{m_W}\right)
\bar{u}_{i}\left[\frac{ \, \sin\alpha}{\sin\beta}\delta_{ij}-
\frac{\sqrt{2} \, \sin(\alpha - \beta)}{g \, \sin\beta}
\left(\frac{m_W}{m_{u_{i}}}\right)(\tilde{Y}_1^u)_{ij}\right]u_{j}H^{0}
\nonumber \\
&  &+ \frac{g}{2}\left(\frac{m_{u_{i}}}{m_W}\right)\bar{u}_{i}
\left[\frac{\cos\alpha}{\sin\beta} \delta_{ij}- \frac{\sqrt{2} \,
\cos(\alpha - \beta)}{g \, \sin\beta}
\left(\frac{m_W}{m_{u_{i}}}\right)(\tilde{Y}_1^u)_{ij}\right]u_{j}
h^{0}
\nonumber \\
& &+ \frac{ig}{2}\left(\frac{m_{u_{i}}}{m_W}\right)\bar{u}_{i}
\left[-\cot\beta \delta_{ij} + \frac{\sqrt{2} }{g \, \sin\beta}
\left(\frac{m_W}{m_{u_{i}}}\right)(\tilde{Y}_1^u)_{ij}\right]
\gamma^{5}u_{j} A^{0}, \label{lageigenstates}
\end{eqnarray}

\noindent where $i=1,2,3$, with $d_{1}=d, d_{2}=s, d_{3}=b,
u_{1}=u, u_{2}=c, u_{3}=t$. The charged leptonic lagrangian is
obtained by substituting down quarks $d_i$ by $l_i$, where
$l_{1}=e,
l_{2}=\mu, l_{3}=\tau$ \\

\par
So, we observe that eq. (\ref{lageigenstates}) includes FCNC
couplings at tree level, which are highly restricted by
experiment. Then, in order to suppress them we consider that the
corresponding Yukawa matrices in eq.(\ref{masa-fermiones}), have
the form of an \textit{hermitian 4-zero texture type}, with a
hierarchy of the form:

\begin{displaymath}
Y_{i}^{f} =  \left( \begin{array}{ccc}
0 & C_{i}^{f} & 0 \\
C_{i}^{f*} & \tilde{B}_{i}^{f} & B_{i}^{f} \\
0 & B_{i}^{f*} & A_{i}^{f}
\end{array}\right).  \qquad \mid A_i^f\mid \, \gg \, \mid \tilde{B}_i^f\mid,\mid B_i^f\mid
,\mid C_i^f\mid.
\end{displaymath}

As we said earlier, it is assumed that each of the Yukawa matrices
has the same form, \textit{hermitian 4-texture ansatz}, which is
inherited to the mass matrix, having

\begin{displaymath}
M_f= \left( \begin{array}{ccc}
0 & C_{f} & 0 \\
C_{f}^{*} & \tilde{B}_{f} & B_{f} \\
0 & B_{f}^{*} & A_{f}
\end{array}\right).\qquad
\mid A^f\mid \, \gg \, \mid \tilde{B}^f\mid,\mid B^f\mid ,\mid
C^f\mid.
\end{displaymath}

Thus, from eq. (\ref{masa-fermiones}) we see that the three
matrices have the same hierarchy
and can be parameterized in the same manner. \\

Accordingly, the $V_{L,R}$ matrices are constructed as the product
of two matrices, one of which contains the complex phases\footnote
{The complete process can be found in Ref. \cite{Fritzsch:1999ee}
and the explicit form of these matrices is given in Refs.
\cite{fourtext} and \cite{primero}.}. Furthermore, as is given in
\cite{primero} we impose the condition $m_{f_{1}}\ll m_{f_{2}},
m_{f_{3}}, \mid A^{f}\mid $ $(f=u, d, l)$, with $A^{f}=
m_{f_{3}}-\beta^{f} m_{f_{2}}$ and $\beta^{f}$ is a number within
the interval $[0,1]$. Therefore, the couplings
$\tilde{Y}_{2}^{d,l}$ and $\tilde{Y}_{1}^{u}$, that appear in eq.
(\ref{lageigenstates}) acquire a simple structure given by:

\begin{eqnarray}
(\tilde{Y}_2^{d,l})_{ij}& = &\frac{\sqrt{m^{d,l}_{i}
m^{d,l}_{j}}}{v}
\tilde{\chi}_{ij}^{d,l}\nonumber  \\
(\tilde{Y}_1^{u})_{ij}& = &\frac{\sqrt{m^{u}_{i} m^{u}_{j}}}{v}
\tilde{\chi}_{ij}^{u}
\end{eqnarray}

Then we keep most of the FCNC processes under control provided
that
$\mid \tilde{\chi}_{ij}^{f} \mid \le O(10^{-1})$.\\

In order to carry out the phenomenological study, we rewrite the
lagrangian (\ref{lageigenstates}) in terms of the parameter of the
model ($\tilde{\chi}_{ij}^{f}$). We display here only the leptonic
part of the Yukawa lagrangian:

\begin{eqnarray}
{\cal{L}}_Y^{l} & = & \frac{g}{2} \bar{l}_{i} \left[\left(
\frac{m_{l_{i}}}{m_W}\right)\frac{\cos\alpha}{\cos\beta}
\delta_{ij} + \frac{\sin(\alpha - \beta)}{\sqrt{2}  \cos\beta}
\left(\frac{\sqrt{m_{l_{i}}
m_{l_{j}}}}{m_W}\right)\tilde{\chi}_{ij}^{l}\right]l_{j}H^{0}
\nonumber \\
& &+ \frac{g}{2} \bar{l}_{i}
\left[-\left(\frac{m_{l_{i}}}{m_W}\right)\frac{\sin\alpha}{\cos\beta}
 \delta_{ij} + \frac{\cos(\alpha - \beta)}{\sqrt{2} \cos\beta}
\left(\frac{\sqrt{m_{l_{i}}
m_{l_{j}}}}{m_W}\right)\tilde{\chi}_{ij}^{l}\right]l_{j} h^{0}
\nonumber \\
& &+ \frac{ig}{2}  \bar{l}_{i}
\left[-\left(\frac{m_{l_{i}}}{m_W}\right)\tan\beta  \delta_{ij} +
\frac{1}{\sqrt{2}  \cos\beta} \left(\frac{\sqrt{m_{l_{i}}
m_{l_{j}}}}{m_W}\right)\tilde{\chi}_{ij}^{l}\right]
\gamma^{5}l_{j} A^{0}. \nonumber
\\
 \label{lagXij}
\end{eqnarray}

We consider that the model parameter is complex in general,
$\tilde{\chi}_{ij}^f=\chi_{ij}^f \exp(\imath\vartheta_{ij})$; with
the real part $\chi_{ij}^f = |\tilde{\chi}_{ij}^f|$ and the effect
of the phase $\vartheta_{ij}$ would be
included as a variation of $-1$ to $1$ on $\chi_{ij}^f$. \\

Having obtained the couplings in terms of the model parameter,
we are ready to calculate the Higgs branching ratios.\\

\section{Branching Ratios of the Neutral Higgs Bosons}

Within the SM, we do not have flavor violation decays at tree
level. The SM width for the decay of the Higgs boson to fermions
at tree level is given by \cite{higgshunter}
\[
\Gamma(\phi^0 \rightarrow\bar{f}f)= \frac{N_{c}}{8\pi} \left(
\frac{gm_{f}}{2m_{W}} \right)^2\beta^{\eta}m_{\phi^{0}},
\]
where $N_c$ is 1 (3) for leptons (quarks). The term in parentheses
stems from the Feynman vertex in the amplitude matrix, while the
kinematic term is given as
$\beta^{2}=1-4m_{f}^{2}/m_{\phi^{0}}^{2}$. For the SM Higgs
$\eta=3$, and when we consider the THDM this value also holds for
$\phi^0=h^0, H^0$, whereas $\eta=1$ for $\phi^0=A^0$. \\

In the case of the THDM-III the decay width for the case of
different fermions includes modified
kinematic factor which involves the three particle masses and is given as follows:\\

\begin{eqnarray}
\Gamma(\phi^{0} \rightarrow\bar{f_i}f_j)& = &
\frac{N_{c}}{8m_{\phi^0}^{3}\pi}\left(\frac{g}{2m_{W}}\right)^2
\xi_{ij}^2 [m_{\phi^0}^2-(m_i+(-)^{n} m_j)^2]
\nonumber\\
& &
\times[(m_i^2-m_j^2-m_{\phi^0}^2)^2-4m_{j}^{2}m_{\phi^0}^2]^{1/2}
\label{Whfifj}
\end{eqnarray}

\noindent where $n$ is even for $h^0, H^0$ and odd for $A^0$.
Thus, we notice that, for the pseudoscalar $A^{0}$, the kinematic
term in equation (\ref{Whfifj}) involves a minus sign. The Yukawa
coupling $\xi_{ij}$, only affects the Higgs-fermion processes
as shown in the Table \ref{tb:couplings}. \\

\begin{table}[hbt]
\renewcommand{\arraystretch}{1.5}
\begin{center}
\begin{tabular}{|l|c|c|c|}
\hline
\textbf{Process}  & \textbf{SM} & \textbf{MSSM} & \textbf{THDM-III} \\
\hline $h^{0}\rightarrow u_i\bar{u}_j$ & $m_{u_i}\delta_{ij}$
   & $\frac{m_{u_{i}}\cos\alpha}{\sin\beta}\delta_{ij}$
   & $[m_{u_{i}}\frac{\cos\alpha}{\sin\beta}\delta_{ij}-
   \frac{\cos(\alpha-\beta)}{\sqrt{2}\sin\beta}\sqrt{m_{u_{i}}m_{u_{j}}}{\tilde{\chi}}_{ij}^u]$ \\
$h^{0}\rightarrow d_i\bar{d}_j$ & $m_{d_i}\delta_{ij}$
   & $\frac{m_{d_i}\sin\alpha}{\cos\beta}\delta_{ij}$
   & $[-m_{d_{i}}\frac{\sin\alpha}{\cos\beta}\delta_{ij}+
   \frac{\cos(\alpha-\beta)}{\sqrt{2}\cos\beta}\sqrt{m_{d_{i}}m_{d_{j}}}\tilde{\chi}_{ij}^d]$ \\
$H^{0}\rightarrow u_i\bar{u}_j$ & - &
$\frac{m_{u_i}\sin\alpha}{\sin\beta}\delta_{ij}$
&$[m_{u_{i}}\frac{\sin\alpha}{\sin\beta}\delta_{ij}-
   \frac{\sin(\alpha-\beta)}{\sqrt{2}\sin\beta}\sqrt{m_{u_{i}}m_{u_{j}}}\tilde{\chi}_{ij}^u]$ \\
$H^{0}\rightarrow d_i\bar{d}_j$ & - &
$\frac{m_{d_{i}}\cos\alpha}{\cos\beta}\delta_{ij}$
  & $[m_{d_{i}}\frac{\cos\alpha}{\cos\beta}\delta_{ij}+
   \frac{\sin(\alpha-\beta)}{\sqrt{2}\cos\beta}\sqrt{m_{d_{i}}m_{d_{j}}}\tilde{\chi}_{ij}^d]$ \\
$A^{0}\rightarrow u_i\bar{u}_j$ & - & $-m_{u_{i}}\cot\beta
\delta_{ij}$
  & $[-m_{u_{i}}\cot\beta\delta_{ij}+
   \frac{\sqrt{m_{u_{i}}m_{u_{j}}}}{\sqrt{2}\sin\beta}\tilde{\chi}_{ij}^u]$ \\
$A^{0}\rightarrow d_i\bar{d}_j$ & - & $-m_{d_{i}}\tan\beta
\delta_{ij}$
  & $[-m_{d_{i}}\tan\beta\delta_{ij}+
   \frac{\sqrt{m_{d_{i}}m_{d_{j}}}}{\sqrt{2}\cos\beta}\tilde{\chi}_{ij}^d]$\\
\hline
\end{tabular}
\renewcommand{\arraystretch}{1.2}
\caption[]{\label{tb:couplings} \it The fermionic vertex of the
neutral Higgs bosons to pair of fermions, $\xi_{ij}$, for
different models. $i,j$ stand for flavors.}
\end{center}
\end{table}

In order to evaluate the Higgs branching ratios, we need to
include the dominant decay modes, in addition to the fermionic
ones. In the next subsection we display these expressions for each
of the Higgs bosons, as taken from
\cite{higgshunter},\cite{Marciano},\cite{Hdecays}.\\

\subsection{Decays of the light Higgs boson, ($h^0$)}

For $h^0$ we shall include the modes $h\rightarrow WW^{*}, ZZ^{*}$
for $m_{h}<2m_{W,Z}$ and $h\rightarrow WW, ZZ$ when kinematically
allowed. The decay width into a real $W$ and virtual $W^*$ boson
is given by \\

\begin{equation}
  \Gamma(h^0\rightarrow
  WW^{*})=4\frac{g^{4}m_{h^{0}}}{512\pi^{3}}\sin^2(\alpha-\beta)F(m_{W}/m_{h^{0}}). \\
\end{equation}
The factor 4 appears because we consider that $W^*\rightarrow tb$
is allowed for
$m_{h}>m_{t}+m_{b}+m_{W}$.\\

While the decay width $h\rightarrow ZZ^{*}$
is given by \\
\begin{eqnarray}
  \Gamma(h^0\rightarrow ZZ^{*})&=&\frac{g^{4}m_{h^{0}}}{2048\pi^{3}}\sin^2(\alpha-\beta)
  [\frac{7-\frac{40}{3}\sin^{2}\theta_{W}+\frac{160}{9}\sin^{4}\theta_{W}}{\cos^{4}\theta_{W}}]
\nonumber \\
&& \nonumber \\
 && \times  F(m_{Z}/m_{h^{0}})
\end{eqnarray}
where\footnote {It should be noted that in references
\cite{higgshunter} and \cite{Marciano} there is a discrepancy of
absolutes values in the formulas. We use the expression given in
Ref. \cite{Marciano}}

\begin{eqnarray}
F(x)&=&-(1-x^{2})(\frac{47}{2}x^{2}-\frac{13}{2}+\frac{1}{x^{2}})-3(1-6x^{2}+4x^{4})\ln(x)
\nonumber\\
&& +3
\frac{1-8x^{2}+20x^{4}}{\sqrt{4x^{2}-1}}\cos^{-1}(\frac{3x^{2}-1}{2x^{3}})
\nonumber
\end{eqnarray}

 Once we reach the $WW$ or $ZZ$ thresholds, we need to consider the decay
widths to the pairs of real vector bosons: \\

\begin{eqnarray}
\Gamma(h^{0}\rightarrow WW,ZZ)&=&\frac{g^{2}m_{h^{0}}^{3}}{k_{h}
64\pi m_{W}^{2}}
  \sin^{2}(\beta-\alpha)\sqrt{1-x_{(W,Z)}}
  \nonumber \\
  & & \times[1-x_{(W,Z)}+\frac{3}{4}x_{(W,Z)}^{2}]
\end{eqnarray}
The factor $k_{h}=1$ for $h^0 \rightarrow WW$ and $k_{h}=2$ for
$h^0
\rightarrow ZZ$; with $x_{(W,Z)}=4m_{W,Z}^{2}/m_{h^{0}}^{2}$. \\

We also take into account the decay mode into gluon pair, which has a decay width:\\

\begin{equation}
\Gamma(h^{0}\rightarrow
gg)=\frac{\alpha_{s}^{2}g^{2}m_{h^{0}}^{3}}
{128\pi^{3}m_{W}^{2}}\frac{\cos^{2}\alpha}{\sin^{2}\beta}|\sum_{q}\tau_{q}[1+(1-\tau_{q})f(\tau_{q})]|^{2}
\end{equation}
where the sum is over all the quarks,
$\tau_{q}=4m_{q}^{2}/m_{\phi^{0}}^{2}$ and
\[
f(\tau_{q})= \left\{ \begin{array}{ll}
  [\sin^{-1}(\sqrt{1/\tau_{q}})]^{2} & if \, \tau_{q}\geq1 \\
  & \\
  \frac{1}{4}[\ln(\eta_{+}/\eta_{-})-i\pi]^{2} & if \, \tau_{q} < 1
\end{array}
\right.
\]
with $\eta_{\pm}\equiv (1 \pm \sqrt{1-\tau_{q}})$. We include only
$top$-contribution in the sum because it is the dominant one,
 since the other quarks have much smaller masses.\\
\par

\subsection{Decays of the heavy $CP-even$ neutral Higgs boson,
($H^0$)}

In this case, the corresponding widths for the decay into a pair
of vector bosons are given by:\\
\begin{eqnarray}
 \Gamma(H^0\rightarrow WW,ZZ)&=&\frac{g^{2}m_{h^{0}}^{3}}{k_{H}64\pi
m_{W}^{2}}\cos^{2}(\beta-\alpha)\sqrt{1-x_{(W,Z)}} \nonumber\\
&&\times [1-x_{(W,Z)}+\frac{3}{4}x_{(W,Z)}^{2}]
\end{eqnarray}
Again, factor $k_{H}=1$ for $H^0 \rightarrow WW$ and $k_{H}=2$ for
$H^0 \rightarrow ZZ$.
 Here also,
$x_{(W,Z)}=4m_{W,Z}^{2}/m_{h^{0}}^{2}$. In fact, we can write some
of the $H^0$ widths using the equations given in section (3.1),
considering the respective mass values for the light and heavy
neutral Higgs bosons. Namely, for $H\rightarrow WW^{*}, ZZ^{*}$, we have: \\
\begin{equation}
\Gamma(H^{0}\rightarrow WW^*,ZZ^*)=\Gamma(h^{0}\rightarrow
WW^*,ZZ^*)\cot^2(\alpha-\beta).
\end{equation}
The expression for the gluonic decay also has the same form as for
$h^0$, but changing the masses and the dependence on the angles
$\alpha-\beta$ of the top-Higgs vertex, namely:
\begin{equation}
\Gamma(H^{0}\rightarrow gg)=\Gamma(h^{0}\rightarrow
gg)\tan^2\alpha.
\end{equation}
Now, in this case we also have the possibility of the Higgs decay
into
$H^0 \rightarrow h^0h^0, A^0A^0$, which have a width given by:\\
\begin{equation}
\Gamma(H^{0}\rightarrow hh)=\frac{g^{2}m_{Z}^{2}f_{h}^{2}}{128\pi
m_{H^{0}}\cos^{2}\theta_{W}}(1-\frac{4m_{h}^{2}}{m_{H^{0}}^{2}})^{1/2}
\end{equation}
where $h=h^0$ or $h=A^0$ and in each case $f_{h}$ consists of the
following mixing angle factors
\[
f_{h}=\left \{ \begin{array}{ll}
  \cos2\alpha\cos(\beta+\alpha)-2\sin2\alpha\sin(\beta+\alpha), & h=h^{0} \\
  \cos2\alpha\cos(\beta+\alpha), & h=A^{0}
\end{array}
\right.
\]

\subsection{Decays of the $CP-odd$ Higgs boson, ($A^0$)}

In the case of the $CP-odd$ Higgs boson, the changes are more evident.
For the decay into gluon pairs we have: \\

\begin{equation}
  \Gamma(A^{0}\rightarrow gg)=\frac{\alpha_{s}^{2}g^{2}m_{A
  ^{0}}^{3}}{128\pi^{3}m_{W}^{2}}
  \cot^{2}\beta |\sum_{i}\tau_{i} f(\tau_{i})|^{2}
\end{equation}
and we also have the possibility of $A^0 \rightarrow Zh^0$. So, we
need to include this mode:\\
\begin{eqnarray}
  \Gamma(A^{0}\rightarrow Zh^{0})&=&\frac{g^{2}\lambda^{1/2}\cos^{2}(\beta-\alpha)}
  {64\pi
  m_{A^{0}}^{3}\cos^{2}\theta_{W}}[m_{Z}^{2}-2(m_{A^{0}}^{2}+
  m_{h^{0}}^{2})+
\nonumber
\\
  && +
  \frac{(m_{A^{0}}^{2}-m_{h^{0}}^{2})^{2}}{m_{Z}^{2}}]
\end{eqnarray}
\vspace{0.5cm}
\noindent with
$\lambda^{1/2}\equiv[(m_{1}^{2}+m_{2}^{2}-m_{3}^{2})-4m_{1}^{2}m_{2}^{2}]^{1/2}$.\\

We also notice that there are no tree level couplings of $A^0$ to
vector boson pairs, as a consequence of assuming $CP$- conservation in the Higgs sector.\\

\section{Results in the THDM-III}

Although, we are working within a toy model we may get some hints
for its possible application or connection to a more fundamental
theory, by performing phenomenological analysis of Higgs decays.
More specifically, we use this model to evaluate the branching
ratios for the three neutral Higgs bosons.

In this THDM, the angles $\alpha$ (the mixing angle in the
$CP-even$ Higgs sector), and $\beta$ (which is the ratio of the
vacuum expectation value of the two doublets), are free
parameters. Unlike the case of the MSSM, where one can fix
$\alpha$ in terms of $\tan\beta$ and $m_{A^0}$. However, we
consider three different scenarios that
depend on how these angles are related to each other:\\
\begin{center}
\begin{tabular}{cll}
 & \bf {scenarios} & \\
\hline
\textbf{1}& \mbox{$\beta-\alpha=\pi/2$} & we obtain for \mbox{$h^0$} the SM-like decays. \\
\textbf{2}&\mbox{$\beta-\alpha=0$} &  we obtain for \mbox{$H^0$} no flavor violation.\\
\textbf{3}&\mbox{$\beta-\alpha=\pi/3$}& we take an angle in
between these extreme cases.\\  \hline
\end{tabular}
\end{center}

\vspace{0.5cm}
 Observing the form of the coefficient, $\xi_{ij}$ for these
scenarios we found that for scenario {\bf 1},  $\xi_{ij}^{h^{0}}$
becomes SM-like and $|\xi_{ij}^{H^{0}}|^2=|\xi_{ij}^{A^{0}}|^2$.
Whereas, for scenario {\bf 2}, $\xi_{ij}^{H^{0}}$ becomes SM-like
and $|\xi_{ij}^{h^{0}}|^2=|\xi_{ij}^{A^{0}}|^2$, as we can see
from Table \ref{tb:scenarios}.

\begin{table}[hbt]
\renewcommand{\arraystretch}{1.5}
\begin{center}
\begin{tabular}{|c|c|c|} \hline
& {\bf $\beta - \alpha = \pi /2$} & $\beta - \alpha = 0$ \\
\hline \hline
  $\xi_{u_{i}u_{j}}^{h^{0}}$ & $m_{u_{i}}\delta_{ij}$
  & $m_{u_{i}}\cot \beta
  \delta_{ij}-\frac{\sqrt{m_{u_{i}}m_{u_{j}}}}{\sqrt{2}\sin\beta}\tilde{\chi}_{ij}^u$ \\
\hline $\xi_{d_{i}d_{j}}^{h^{0}}$ & $m_{d_{i}}\delta_{ij}$
  & $-m_{d_{i}}\tan \beta
  \delta_{ij}+\frac{\sqrt{m_{d_{i}}m_{d_{j}}}}{\sqrt{2}\cos\beta}\tilde{\chi}_{ij}^d $\\
\hline $\xi_{u_{i}u_{j}}^{H^{0}}$
  &$-m_{u_{i}}\cot \beta
  \delta_{ij}+\frac{\sqrt{m_{u_{i}}m_{u_{j}}}}{\sqrt{2}\sin\beta}\tilde{\chi}_{ij}^u$
  & $m_{u_{i}}\delta_{ij}$ \\
\hline $\xi_{d_{i}d_{j}}^{H^{0}}$
  & $m_{d_{i}}\tan \beta
  \delta_{ij}-\frac{\sqrt{m_{d_{i}}m_{d_{j}}}}{\sqrt{2}\cos\beta}\tilde{\chi}_{ij}^d$
  & $m_{d_{i}}\delta_{ij}$ \\
\hline
  $\xi_{u_{i}u_{j}}^{A^{0}}$
  & $-m_{u_{i}}\cot \beta
  \delta_{ij}+\frac{\sqrt{m_{u_{i}}m_{u_{j}}}}{\sqrt{2}\sin\beta}\tilde{\chi}_{ij}^u$
  & $-m_{u_{i}}\cot \beta
  \delta_{ij}+\frac{\sqrt{m_{u_{i}}m_{u_{j}}}}{\sqrt{2}\sin\beta}\tilde{\chi}_{ij}^u $\\
\hline
  $\xi_{d_{i}d_{j}}^{A^{0}}$
  & $-m_{d_{i}}\tan \beta
  \delta_{ij}+\frac{\sqrt{m_{d_{i}}m_{d_{j}}}}{\sqrt{2}\cos\beta}\tilde{\chi}_{ij}^d$
  & $-m_{d_{i}}\tan \beta
  \delta_{ij}+\frac{\sqrt{m_{d_{i}}m_{d_{j}}}}{\sqrt{2}\cos\beta}\tilde{\chi}_{ij}^d$\\
\hline
\end{tabular}
\renewcommand{\arraystretch}{1.2}
\caption[]{\label{tb:scenarios} \it Explicit form of the
coefficients $\xi_{ij}$ for the two extreme scenarios, {\bf 1} and
{\bf 2}.}
\end{center}
\end{table}

\begin{figure}[hbt]
\begin{center}
\mbox {\includegraphics[width=10cm, height=8cm]{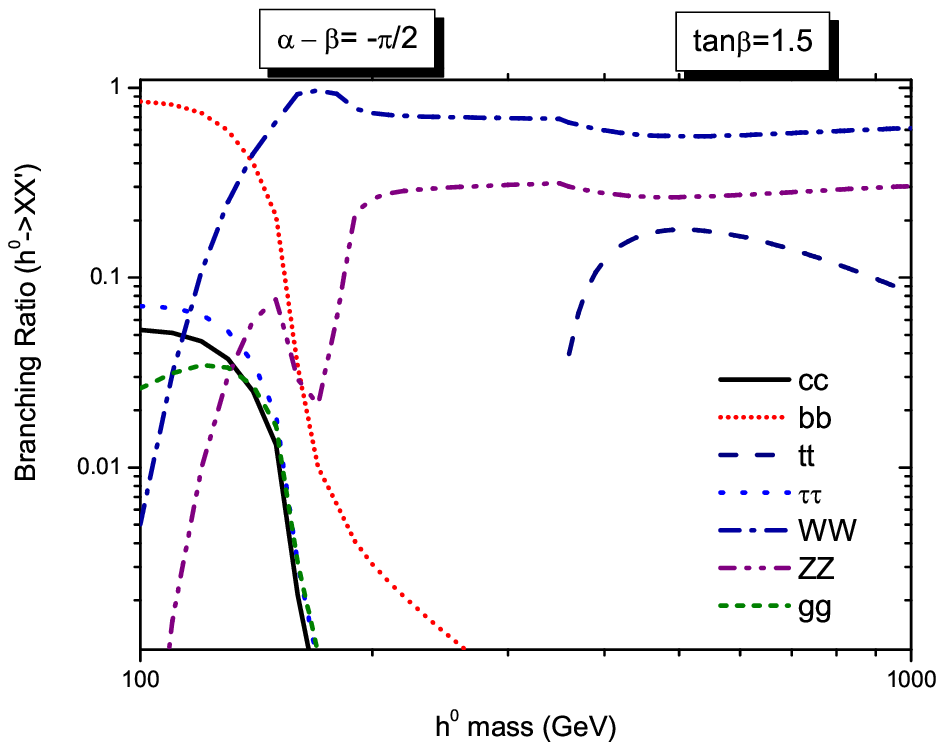 }}
\caption[]{\label{fg:SMH} \it Branching ratios for $h^0$ to pair
of fermions in the scenario \textbf{1}, taking $\tan\beta=1.5$.}
\end{center}
\end{figure}

\subsection{Light Higgs Boson ($h^0$)}

In order to constrain the value of the model parameter,
$\chi_{ij}$, we consider the contributions that the model may
induce for the specific decay $h^0\rightarrow b\bar{b}$, see Table
\ref{tb:bb}. We let the parameter $\chi_{ij}$ vary from $-1$ to
$1$. We notice that for $\chi=-0.2$, the corrections to
$h^0\rightarrow b\bar{b}$ are minimal. One can see that the
largest difference from MSSM-like couplings, {\it i.e.}
$\chi_{ij}=0$, occurs at $\chi=1$.

\begin{table}[hbt]
\begin{center}
\small{
\begin{tabular}{|c|c|c|c|c|c|c|} \hline
  &  $tan\beta=5$ & $tan\beta=10$ & $tan\beta=15$ & $tan\beta=20$  & $tan\beta=30$&
  $tan\beta=50$\\
  \hline
$\chi_{ij}$=1.0& 0.63958& 0.69383& 0.70424& 0.70793& 0.71059&
0.71196\\
$\chi_{ij}$=0.5& 0.88958& 0.90235& 0.90474& 0.90558&  0.90618&
0.90649\\
$\chi_{ij}$=0.0& 0.91325& 0.91848& 0.91944& 0.91978&  0.92002&
0.92014\\
$\chi_{ij}$=-0.2&  0.91404& 0.91815& 0.91891& 0.91917&
0.91936& 0.91945\\
$\chi_{ij}$=-0.5&    0.91294& 0.91607& 0.91664& 0.91684& 0.91698& 0.91705\\
$\chi_{ij}$=-1.0&    0.90933& 0.91165& 0.91207&
0.91221& 0.91231& 0.91237\\
\hline
\end{tabular}}
\renewcommand{\arraystretch}{1.2}
\caption[]{\label{tb:bb} \it Branching ratios for $h^0$ to a pair
of b-quarks for different values of the parameter $\chi_{ij}$ and
$tan\beta$. Observe that we obtain the MSSM decay in the value of
$\chi_{ij}=0$.}
\end{center}
\end{table}

As we mentioned, for the first scenario, where $\alpha - \beta =
-\pi/2$, there is no flavor violation, so, by taking
$\tan\beta=1.5$, we see in Fig. \ref{fg:SMH}, that the SM
branching ratios are reproduced
\cite{HSM},\cite{PDG}.\\

\begin{figure}[hbt]
\begin{center}
\mbox {\includegraphics[width=13cm, height=7cm]{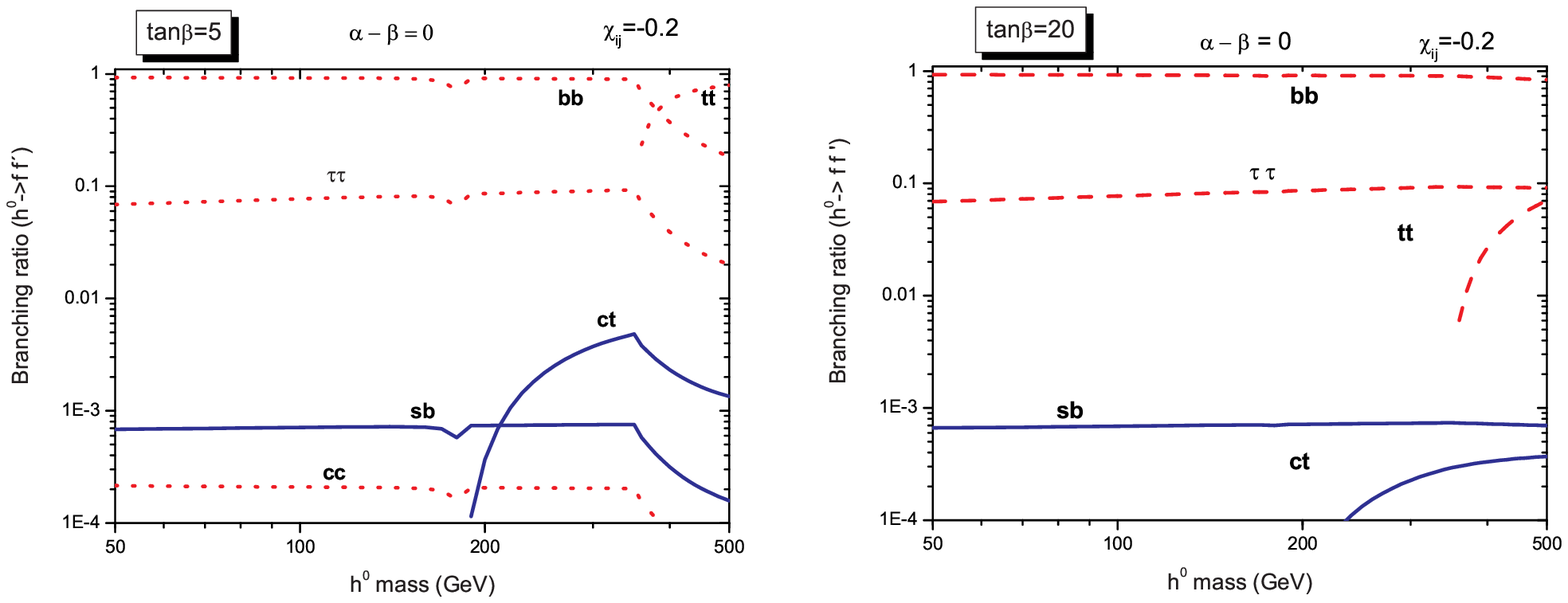}}
\caption[]{\label{fg:HP0}\it Branching ratios for $h^0$ to pair of
fermions, where $\beta=\alpha$ for two values of $\tan\beta=5,20$
and $\chi_{ij}=-0.2$.}
\end{center}
\end{figure}

\begin{figure}[hbt!]
\begin{center}
\mbox {\includegraphics[width=13cm, height=7cm]{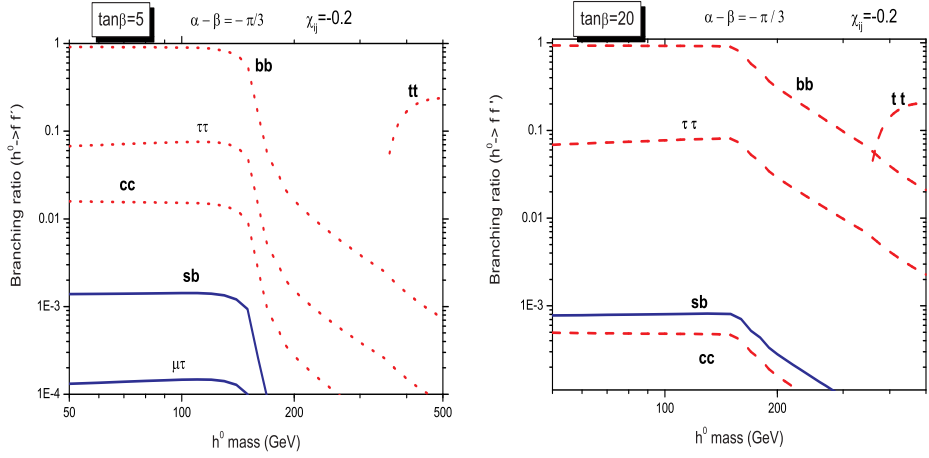}}
\caption[]{\label{fg:HP3}\it Branching ratios for $h^0$ to pair of
fermions, where $\beta-\alpha=\pi/3$ for two values of
$\tan\beta$, 5 and 20; with $\chi_{ij}=-0.2$.}
\end{center}
\end{figure}

In Figs. \ref{fg:HP0} and \ref{fg:HP3} the dependence of the
branching ratios on the light Higgs mass $h^0$ are shown, for
$\alpha = \beta$ and $\alpha - \beta = -\pi/3$; taking
$\chi_{ij}=-0.2$, for $\tan\beta=5$ and $\tan\beta=20$. These
representative values are taken because, the former case, is a
small value {\it i.e.} $1\leq tan\beta \leq 10$, here the
dependence of the branching ratios vary  more strongly. In the
letter case: large $\tan\beta$, the value is chosen where the
behavior of the branching ratios becomes, in general, quite
independent of $\tan\beta$, for different channels of the light Higgs boson.\\

\begin{figure}[hbt]
\begin{center}
\mbox {\includegraphics[width=13cm, height=7cm]{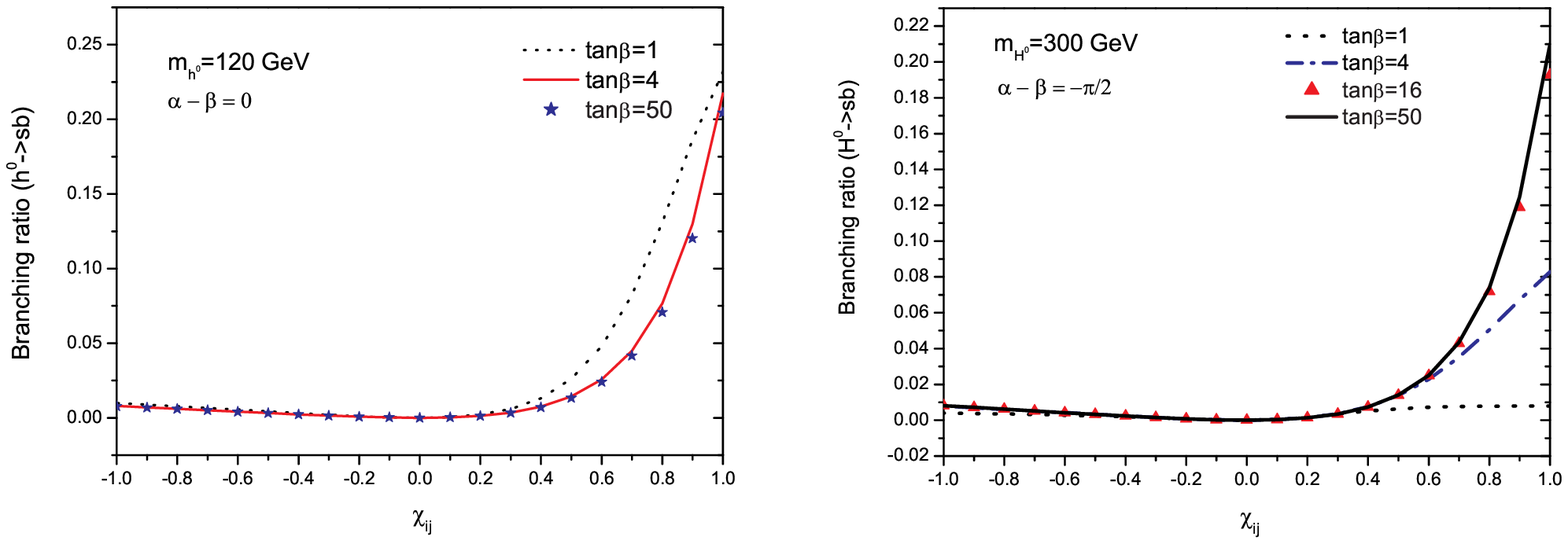}}
\caption[]{\label{fg:HPGsb}\it Dependence of the sb decay channel
of $h^0$ and $H^0$ on $\chi_{ij}$ for different values of
$\tan\beta$, where the Higgs mass is $m_{h^0}=120 Gev$ in scenario
{\bf 2}.}
\end{center}
\end{figure}

In order to achieve maximal flavor violation for the decay $h^0
\rightarrow sb$, the parameter space must corresponds to
$\chi_{ij}\rightarrow 1$, and in this scenario the dependence on
$\tan\beta$ is very mild, even though for small values of
$\tan\beta$ the flavor violation is enhanced. If these flavor
violation decay is highly restricted by the experiment, the
parameter space should be in the region where $ \chi_{ij}$ is
close to zero, as can be seen in the first graph of figure
\ref{fg:HPGsb}.
 We are setting the Higgs mass value at $120 GeV$ and considering
the best scenario for this decay, $\beta=\alpha$.\\
A summary of maximal FV modes is shown in Table \ref{tb:fvh0}.

\begin{table}[hbt!]
\begin{center}
\begin{tabular}{|c|c|c|c|c|}
\hline
& {\bf $BR(h^0\rightarrow ff')$} & {\bf $\tan\beta$} & {\bf $m_{h^0}$} & {\bf scenario}\\
\hline
  $ct-channel$ & $\sim 10^{-3}$ & 5 & $\sim 220-500$ & 2\\
  $sb-channel$ & $\sim 10^{-3}$ & 5 & $\sim  50-140$ & 3\\
  $\mu\tau-channel$ & $\sim 10^{-4}$ & 5 & $\sim 50-150$ & 3\\
  \hline
\end{tabular}
\caption[]{\label{tb:fvh0} \it  Maximal BR for the flavor
violating decays of the light Higgs boson, ($h^0$) for
$\chi_{ij}=-0.2$.}
\end{center}
\end{table}

\subsection{Heavy neutral Higgs bosons ($H^0$)}

\begin{figure}[hbt!]
\begin{center}
\mbox {\includegraphics[width=13cm, height=8cm]{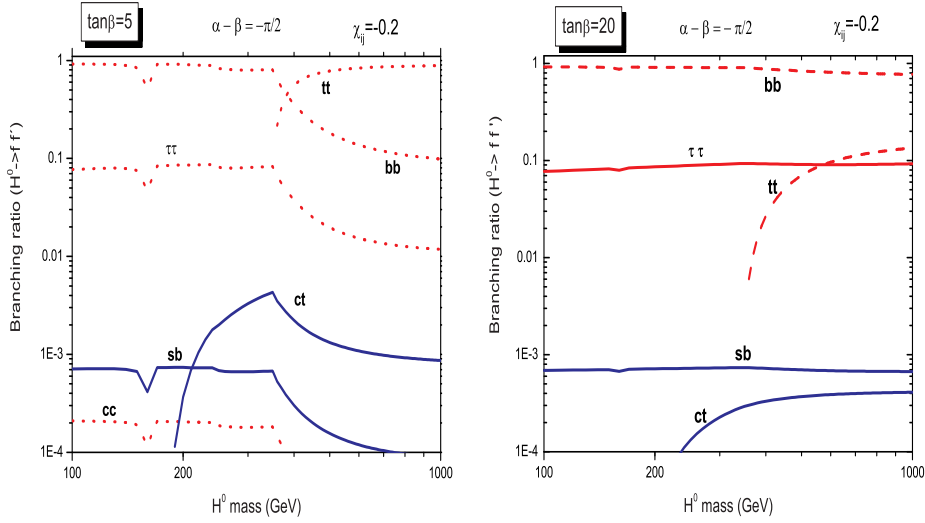}}
\caption[]{\label{fg:HG2}\it Branching ratios for $H^0$ to pair of
fermions, where $\beta-\alpha=\pi/2$ for two values of
$\tan\beta=5,20$ and $\chi_{ij}=-0.2$.}
\end{center}
\end{figure}

\begin{figure}[hbt!]
\begin{center}
\mbox {\includegraphics[width=13cm, height=8cm]{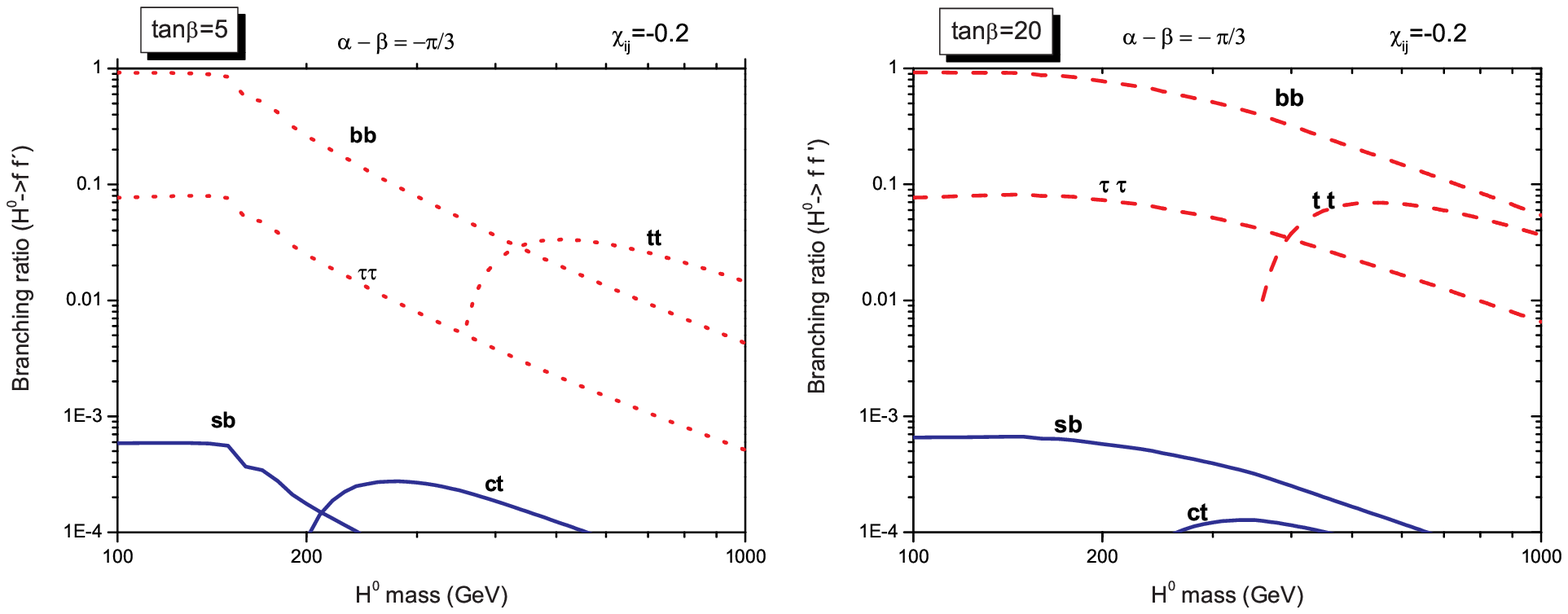}}
\caption[]{\label{fg:HG3}\it Branching ratios for $H^0$ to pair of
fermions, where $\beta-\alpha=\pi/3$ for two values of
$\tan\beta=5,20$ and $\chi_{ij}=-0.2$.}
\end{center}
\end{figure}

Because for $H^0$, the flavor violation couplings vanish for
$\alpha -\beta = 0$, we only show results for scenarios {\bf 1}
and {\bf 3}, for $\tan\beta=5$ and $\tan\beta=20$, displayed on
Figs. \ref{fg:HG2} and \ref{fg:HG3}, respectively. In this case,
the vector boson  decay channel is open for almost the entire mass
range, in fact it is the main decay mode, leaving the fermionic
modes as a minor contribution of the decay rate. However, there is
a region where the fermionic decay becomes important, when the top-threshold is reached.\\

 In this case we also explore the parameter space to visualize the
dependance of the flavor violation decay $H^0 \rightarrow sb$ on
the parameters. In this case, there is a strong dependence, mostly
on large values of $\tan\beta$ and a model parameter close to 1: $
\tan\beta
> 10$ and $\chi_{ij}\rightarrow 1$. On the other hand, the flavor
violating decays are reduced in the region of parameter space
where $\tan\beta$ is small and $\chi_{ij}$ is close to zero as can
be seen in the second graph of figure \ref{fg:HPGsb}. Here we have
fixed the heavy Higgs mass at $300 GeV$ and taking the more
favored scenario for the enhanced of these decays, $\beta-\alpha=\pi/2$.
A summary of maximal FV modes for $H^0$ is shown in Table \ref{tb:fvHG0}.\\

\begin{table}[hbt!]
\begin{center}
\begin{tabular}{|c|c|c|c|c|}
\hline
        & {\bf $BR(H^0\rightarrow ff')$} & {\bf $\tan\beta$} & {\bf $m_{H^0}$} & {\bf scenario} \\
        \hline
  $ct-channel$ & $\sim 10^{-3}$ & 5 & $\sim 220-600$ & 1\\
  $sb-channel$ & $\sim 10^{-4}$ & 5,20 & $\sim  50-650$ & 1\\
  $\mu\tau-channel$ & $\sim 10^{-5}$ & 5,20 & $\sim 50-1000$ & 1\\
  \hline
\end{tabular}
\caption[]{\label{tb:fvHG0} \it Maximal flavor violating decays
for the light Higgs boson, $H^0$ for $\chi_{ij}=-0.2$.}
\end{center}
\end{table}

\subsection{Heavy neutral CP-odd Higgs boson ($A^0$)}

For the pseudoscalar $A^0$, the fact that there is no coupling to
gauge bosons makes the flavor violating signal more stable. The
fermionic final states are the main decays even for the large mass
range, (since we are not in the context of the MSSM, no possible
decays into sparticles are take into account). Another important
issue about the pseudoscalar is that its fermion couplings do not
depend on the mixing angle $\alpha$, making its branching ratio
almost independent of the chosen scenario, nevertheless there is a
slightly difference
coming from the width $\Gamma(A^{0}\rightarrow Zh^{0})$.\\

\begin{figure}[hbt]
\begin{center}
\mbox {\includegraphics[width=13cm, height=8cm]{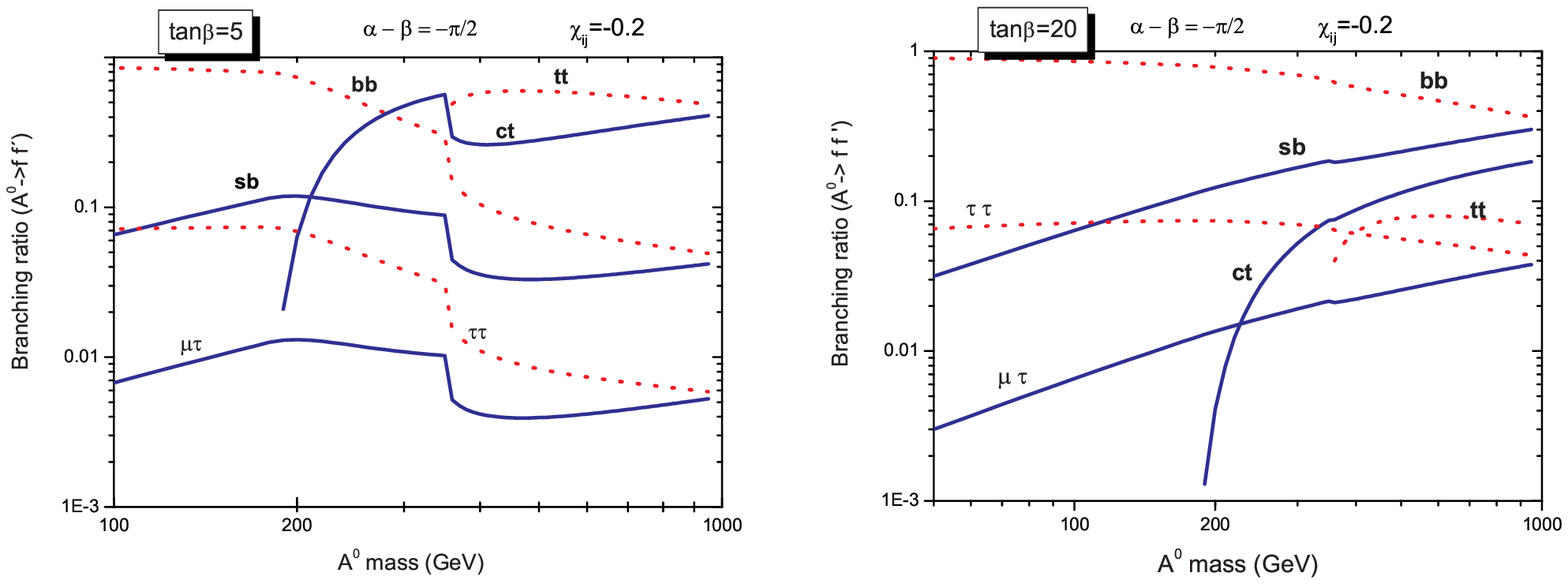}}
\caption[]{\label{fg:HA2}\it Branching ratios for $A^0$ to pair of
fermions, where $\beta-\alpha = \pi/2$ for two values of
$\tan\beta=5,20$ and $\chi_{ij}=-0.2$.}
\end{center}
\end{figure}

\begin{figure}[hbt!]
\begin{center}
\mbox {\includegraphics[width=13cm, height=8cm]{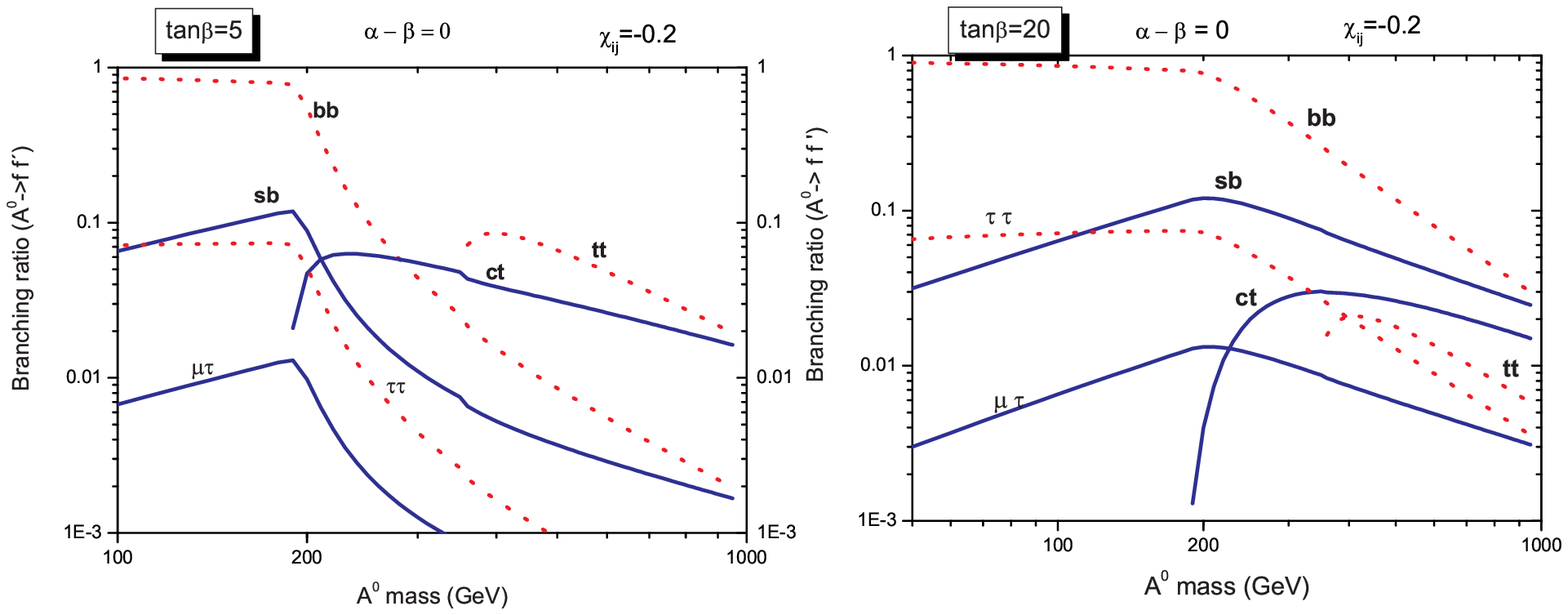}}
\caption[]{\label{fg:HA0}\it Branching ratios for $A^0$ to pair of
fermions, where $\beta=\alpha$ for two values of $\tan\beta=5,20$
and $\chi_{ij}=-0.2$.}
\end{center}
\end{figure}

\begin{figure}[hbt!]
\begin{center}
\mbox {\includegraphics[width=13cm, height=8cm]{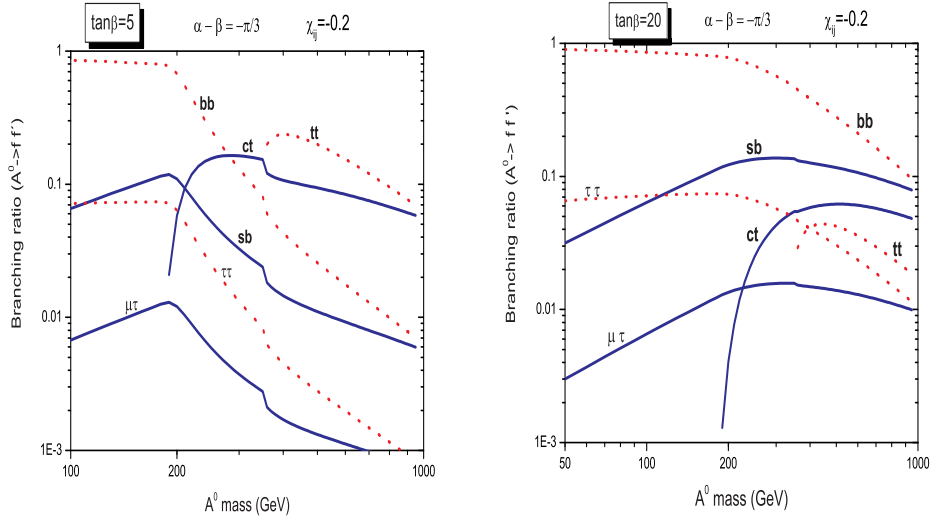}}
\caption[]{\label{fg:HA3}\it Branching ratios for $A^0$ to fermion
pairs, where $\beta-\alpha = \pi/3$ for two values of
$\tan\beta=5,20$ and $\chi_{ij}=-0.2$.}
\end{center}
\end{figure}

By fixing the two values of $\tan\beta$, as done before, and for
the three scenarios:  $\beta-\alpha = \pi/2$, $\beta = \alpha $
and $\beta-\alpha = \pi/3$ we obtain what is depicted on Figs.
\ref{fg:HA2}, \ref{fg:HA0} and \ref{fg:HA3}. As inferred from
these figures, the signals for flavor violating decays would in
general be clearer and higher for $A^0$. In particular, the
branching ratio $A^{0}\rightarrow ct$ is very high in the mass
range $290 < m_{A^0}< 2m_t$, where it becomes the dominant decay
mode, reaching a branching ratio larger than about $50\%$, as
shown in Fig. \ref{fg:HA2}.\\

\begin{figure}[hbt!]
\begin{center}
\mbox {\includegraphics[width=13cm, height=7cm]{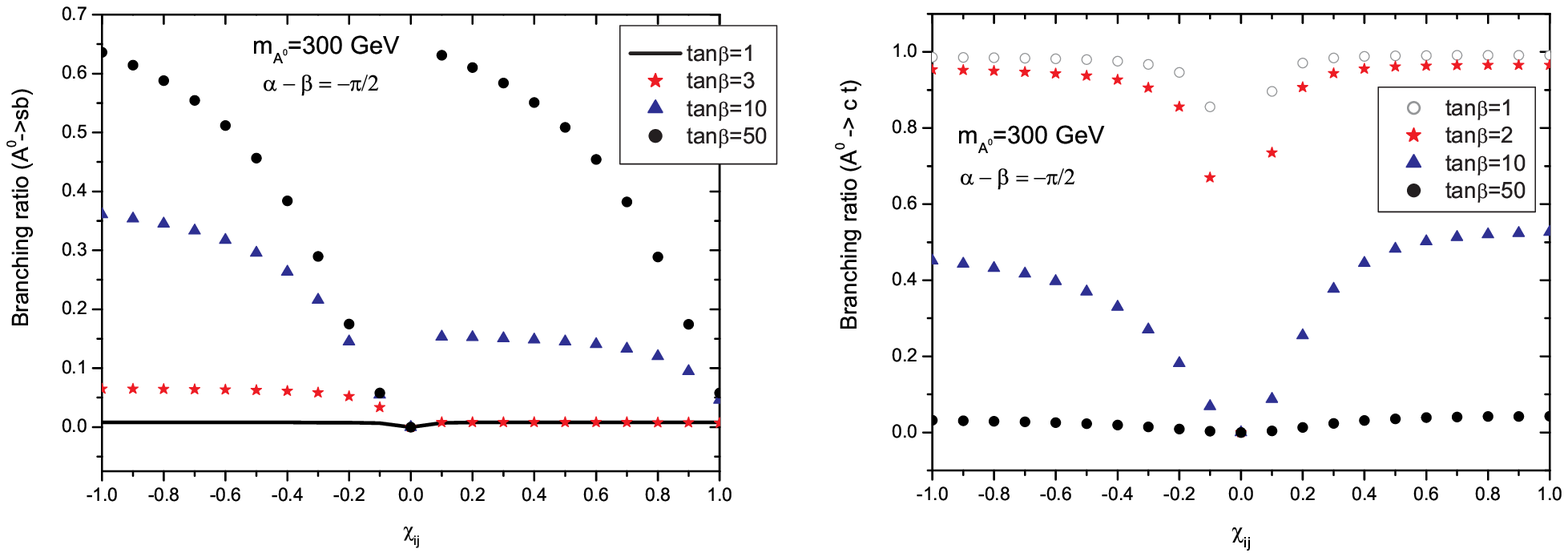}}
\caption[]{\label{fg:HAsbct}\it Dependence of the decay
$A^0\rightarrow sb,ct$ on $\chi_{ij}$ for different values of
$\tan\beta$, where $m_{A^0}=300 GeV$, in scenario \bf 1.}
\end{center}
\end{figure}

In order to explore the dependence of the flavor violating decays
for $A^0$ on the parameter space, we plot in figure
\ref{fg:HAsbct} the branching ratio $A^0\rightarrow sb$ and
$A^0\rightarrow ct$ as function of $\chi_{ij}$ for different
values of $\tan\beta$, obtaining. In left plot of Fig.
\ref{fg:HAsbct}, the regions with large $\tan\beta$ and with
$\chi_{ij}\sim -1$ or $\chi_{ij}\sim 0$ (with $\chi_{ij}>0$), have
a large branching ratio for $A^0\rightarrow sb$; while for the
decay $A^0\rightarrow ct$, as shown on right plot of Fig.
\ref{fg:HAsbct}, the two regions correspond to small $\tan\beta
\sim 1$ and $\chi_{ij}\sim \pm 1$. A summary of maximal FV modes
for $A^0$ is shown in Table \ref{tb:fvA0}.

\begin{table}[hbt!]
\begin{center}
\begin{tabular}{|c|c|c|c|c|}
\hline
& {\bf $BR(A^0\rightarrow ff')$} & {\bf $\tan\beta$} & {\bf $m_{A^0}$} & {\bf scenario}\\
\hline
  $ct-channel$ & $\sim 0.5$ & 5 & $\sim 210-350$ & 1\\
  $sb-channel$ & $\sim 10^{-1}$ & 20 & $> 950$ & 1\\
  $\mu\tau-channel$ & $\sim 0.03$ & 20 & $> 650$ & 1\\
  \hline
\end{tabular}
\caption[]{\label{tb:fvA0} \it Maximal flavor violating decays for
the pseudoscalar Higgs boson, $A^0$ for $\chi_{ij}=-0.2$.}
\end{center}
\end{table}

\section{Conclusions}

The THDM-III, enable us to study, in a more general way, the
effects of the Yukawa couplings. In particular, the flavor
violating decays could be restricted at low energies by the
specific form of {\it 4-zero textures}. In such case, the Yukawa
matrices give the correct mass spectrum and mixing angles to the
fermion sector. Using this model \cite{primero}, we explore the
branching ratios at tree level of the neutral Higgs sector over
the whole range of the neutral Higgs masses. The model parameter
is restricted in order to keep the flavor conserving decay modes
near the MSSM case, and to avoid FCNC at low energies, though
having in mind that as long as no experimental evidence is
produced the parameter
space is only mildly restricted.\\

For the specific parameter values $\chi$ $(=-0.2)$ and
$\tan\beta$, we have obtained maximal flavor changing decays at
tree level for the $h^0$ of the order of $10^{-3}$ for
$ct-channel$ in scenario {\bf 2} and similar results for
$sb-channel$ in scenario {\bf 3}, values of the order of $10^{-4}$
were obtained for $\mu\tau-channels$ in scenario {\bf 3}, for
$\tan\beta=5$ in all three cases. The values of these branching
ratios decays are reduced for $\tan\beta=20$ down to the order of
$10^{-4}$ for the quark channels, end even
lower values for lepton violation decay, which are reduce to about $10^{-5}$.\\

In the case of $H^0$ we obtained maximal flavor changing decays of
the order of $10^{-3}$ for $ct-channel$, $10^{-4}$ for
$sb-channel$, and $10^{-5}$ for $\mu\tau-channel$, all cases in
scenario {\bf 1} and
$\tan\beta=5$.\\

Finally, for the pseudoscalar $A^0$, we have obtained significant
flavor violating decays rates (as large as $50\%$) for the
$A^0\rightarrow ct$ and as high as $30\%$ for $A^0\rightarrow sb$,
for specific values of the parameter space, particularly in
scenario {\bf 1}. While the former decay is enhanced at
$\tan\beta=5$, the other two flavor violation decays, $sb$ and
$\mu\tau$, are larger in the case of $\tan\beta=20$ and $m_{A^0}\sim 1TeV$.\\

From this analysis we also conclude that the value for the
parameter space $\chi_{ij}=-0.2$, does not lead to large flavor
violating decay rates in the cases of the two $CP-even$ Higgs
bosons; while, in the other hand, increases  flavor violating
decay rates for the pseudoscalar $A^0$.\\

We have explored the complete parameter space of this model in
order to determinate the areas where $h^0, H^0, A^0$ reach maximal
branching ratios. Studying these modes at future colliders (LHC)
could be important to find new Higgs signals and explore the
origin of flavor
\cite{Diaz-Cruz:1998qc}.\\

\vspace*{0.5cm}
\noindent{\bf Acknowledgments.}\\

\noindent M. G.-B. and  R. N.-P. gratefully acknowledge the
provision of funds provided by CONACyT (M\'exico) as Ph. D.
fellowships. We also thank Dr. L. D\'{\i}az-Cruz and Dr. A. Rosado
for suggesting this problem and for fruitful discussions. We are
also thankful to Dr. A. Flores-Riveros for the writing review of
this paper. M.G.-B, thanks the ICTP for financial support, which
enabled her attendance to the {\it Summer School on Particle
Physics 2005}, held at Trieste, Italy, where part of this work was
done.


\begin{thebibliography}{99}

\bibitem{FN}
C. D. Frogatt and H. B. Nielsen, Nucl. Phys. B{\bf 147}, 277
(1979).

\bibitem{cheng-li} For example see:
T. P. Cheng and  L. F Li, {\it Gauge theory of elementary particle
physics.} (Oxford University Press, 1984)

\bibitem{Fritzsch:2001py}
  H.~Fritzsch,
  arXiv:hep-ph/0111051.

\bibitem{glashow}
S.Glashow and S. Weinberg, Phys. Rev.D {\bf 15}, 1958 (1977)

\bibitem{higgshunter} For a review see:
J. F Gunion, H. E Haber, G. L. Kane, and S. Dawson, {\it The Higgs
Hunter's Guide} (Addison-Wesley, Reading, MA, 1990)

\bibitem{Martin:1997ns} For a review see:
  S.~P.~Martin,
  arXiv:hep-ph/9709356.

\bibitem{Savage:1991qh}
  See for example: M.~J.~Savage,
  Phys.\ Lett.\ B {\bf 266}, 135 (1991).

\bibitem{Fritzsch:1999ee}
  H.~Fritzsch and Z.~z.~Xing,
  Prog.\ Part.\ Nucl.\ Phys.\  {\bf 45} (2000) 1
  [arXiv:hep-ph/9912358].

\bibitem{Cheng:1987rs}
T.~P.~Cheng and M.~Sher,
Phys.\ Rev.\ D {\bf 35}, 3484 (1987).

\bibitem{ansatz}See for example:
  D.~Atwood, L.~Reina and A.~Soni,
  Phys.\ Rev.\ D {\bf 55} (1997) 3156
  [arXiv:hep-ph/9609279].
  W.~S.~Hou,
  Phys.\ Lett.\ B {\bf 296} (1992) 179.
  M.~E.~Luke and M.~J.~Savage,
  Phys.\ Lett.\ B {\bf 307} (1993) 387
  [arXiv:hep-ph/9303249].

\bibitem{fourtext} H. Fritzsch and Z. Z. Xing,
  Phys.\ Lett.\ B {\bf 555}, 63 (2003)
  [arXiv:hep-ph/0212195].

\bibitem{Xing}
  Z.~z.~Xing and H.~Zhang,
  J.\ Phys.\ G {\bf 30}, 129 (2004)
  [arXiv:hep-ph/0309112].

\bibitem{primero} J.~L.~Diaz-Cruz, R.~Noriega-Papaqui and A.~Rosado,
  Phys.\ Rev.\ D {\bf 69}, 095002 (2004)
  [arXiv:hep-ph/0401194].

\bibitem{segundo} J.~L.~Diaz-Cruz, R.~Noriega-Papaqui and A.~Rosado,
  Phys.\ Rev.\ D {\bf 71}, 015014 (2005)
  [arXiv:hep-ph/0410391].

\bibitem{Alexis}
  R.~Martinez, J.~A.~Rodriquez and D.~A.~Milanes,
  Phys.\ Rev.\ D {\bf 72} (2005) 035017
  [arXiv:hep-ph/0502087].

\bibitem{Marciano} W. Y. Keung and W. J. Marciano, Phys.Rev. D
{\bf 30}, 248 (1984);

\bibitem{Hdecays}
E.~Barradas, J.~L.~Diaz-Cruz, A.~Gutierrez and A.~Rosado,
Phys.\ Rev.\ D {\bf 53}, 1678 (1996);
J.~L.~Diaz-Cruz and M.~A.~Perez,
Phys.\ Rev.\ D {\bf 33}, 273 (1986).

\bibitem{HSM}
A.~Djouadi, J.~Kalinowski and M.~Spira,
Comput.\ Phys.\ Commun.\  {\bf 108}, 56 (1998)
[arXiv:hep-ph/9704448];
 E.~Accomando {\it et al.} [ECFA/DESY LC
Physics Working Group Collaboration],
Phys.\ Rept.\  {\bf 299}, 1 (1998) [arXiv:hep-ph/9705442].

\bibitem{PDG} Particle Data Group. Phys. Lett. B {\bf 592}, 1
(2004). (URL:http://pdg.lbl.gov)

\bibitem{Diaz-Cruz:1998qc}
J.~L.~Diaz-Cruz, H.~J.~He, T.~Tait and C.~P.~Yuan,
Phys.\ Rev.\ Lett.\  {\bf 80}, 4641 (1998) [arXiv:hep-ph/9802294];
C.~Balazs, J.~L.~Diaz-Cruz, H.~J.~He, T.~Tait and C.~P.~Yuan,
Phys.\ Rev.\ D {\bf 59}, 055016 (1999) [arXiv:hep-ph/9807349];

J.~L.~Diaz-Cruz and J.~J.~Toscano,
Phys.\ Rev.\ D {\bf 62}, 116005 (2000) [arXiv:hep-ph/9910233].

\end{thebibliography}
\end{document}